\documentclass[twoside]{elsart}
\usepackage{epsfig}%
\usepackage{graphicx}
\usepackage{dcolumn}
\usepackage{bm}
\usepackage{amsmath,amssymb}
\begin{document}
\begin{frontmatter}
\title{Aging in financial market}\author[usa]{Simone Bianco}\footnote{{\it
  Corresponding author:} sbianco@unt.edu}, \author[usa,it1,it2]{Paolo
  Grigolini}
\address[usa]{Center for Nonlinear Science, University of North Texas,
  P.O. Box 311427,Denton, Texas 76203-1427, USA}
\address[it1]{Istituto dei Processi Chimico Fisici del CNR, Area della
  Ricerca di Pisa, Via G. Moruzzi, 56124, Pisa, Italy}
\address[it2]{Dipartimento di Fisica "E.Fermi" - Universit\`a di Pisa, Largo  Pontecorvo, 3 56127 PISA}
\date{\today}
\begin{abstract}
We analyze the data of the Italian and U.S. futures on the stock markets and
we test the validity of the Continuous Time Random Walk assumption for the survival probability of
the returns time series via a renewal aging experiment. We also study the
survival probability  of returns sign and apply a coarse graining procedure to
reveal the renewal aspects of the process underlying its dynamics.  
\end{abstract}
\begin{keyword}
Waiting-time; subordination; continuous time random walk; statistical finance;
\\ {{\it PACS: \ }} 05.40.-a, 89.65.Gh, 02.50.Cw, 05.60.-k, 47.55.Mh
\end{keyword}
\end{frontmatter}

\section{Introduction}
In the last few years there has been an increasing interest for the subject of fractional derivatives and for the physical, biological, sociological and econophysical application of this theoretical tool. We refer the reader to the recent mini-review 
of Scalas~\cite{fractional}, who addresses this issue with a special emphasis on the econophysical application.  There is close connection between fractional calculus and Continuous Time Random Walk (CTRW) perspective~\cite{ctrw}, and consequently with the subject of non-Markov master equations~\cite{kenkre}. However, these connections, and the formal equivalence of the equations motion for the probability time evolution as well, do not necessarily imply physical equivalence. For instance, the generalized diffusion equation discussed in Refs.~\cite{balescu,rasit}, in the super-diffusional case is incompatible 
with the CTRW perspective, insofar as there are no renewal events, resetting to zero the system's memory.

In the case of financial market, there have been recently several
contributions, related to some different extents, to the subject of fractional
derivatives and CTRW~\cite{fin1,fin2,fin3,fin4,fin5,fin6,fin7}. Of special interest for
this paper is the fact that the authors of  Ref.~\cite{fin1} using fractional
calculus made the theoretical prediction that the high-frequency financial
market data generate a Survival Probability (SP) with the form of a Mittag-Leffler function. The later paper of Ref.~\cite{fin2}, an empirical analysis,  confirmed the theoretical prediction of Ref.~\cite{fin1}, even though the restriction to short times forced the authors of Ref.~\cite{fin2} to approximate the Mittag-Leffler SP with a stretched exponential.

In this paper we show how to go beyond this limit, so as to reveal the Mittag-Leffler SP in its entirety. To realize this purpose, we study the SP for two different financial markets, the
futures on the Italian stock index, S$\&$P MIB, and the futures on the US
S$\&$P500.  We discuss the emergence of a Mittag-Leffler SP  from the returns time series at our disposal. However, in principle,
the mere analysis of the form of the SP is not yet enough to ensure that the dynamics of financial market are compatible with the CTRW perspective. This is so because, as earlier pointed out, the formal equivalence between two generalized master equations, one of Liouville origin~\cite{grigo} and the other of CTRW origin, does not ensure the full dynamic equivalence (see also~\cite{rasit}). 

In the last few years it became clear that CTRW yields renewal aging~\cite{barkaiaging}, and the authors of Refs.~\cite{brok,paradisi} have proposed numerical experiments to assess this important property in physical systems. We plan to use the prescription of Ref.~\cite{paradisi} as an efficient method to establish the renewal nature of the financial market process. Proving renewal aging is equivalent to ruling out the possibility that the generalized master equation generating the Mittag-Leffler SP  might have a Liouville origin, with trajectory memory and no renewal event. It has to pointed out, however, that the aging experiment must be done with caution. In fact, as shown in Ref.~\cite{paradisi}, the renewal event might be masked by a cloud of secondary events, of Poisson nature, generating the wrong impression that the process is not renewal, and that its memory is a property of the individual trajectories. The authors of Ref.~\cite{memory} denoted these producing camouflage events as \emph{pseudo events}, and have established that in the case of heart beating analysis they are generated by the data processing procedure itself used for the statistical analysis. In other cases, the pseudo events can be triggered by the renewal events: for instance, the authors of Ref.~\cite{mega}, studied the seismic fluctuations in South California and made the conjecture that the main shocks are renewal events, triggering the Omori after shocks, which are an example of pseudo events in this case. The main shocks are not necessarily the seismic fluctuations of large intensity, and the seismic fluctuation caused by them might be larger, thereby making the renewal events invisible. \\

How to detect the renewal events? This is a challenging issue that has been mainly addressed  so far through the scaling analysis. There are reasons to believe that when the time series is converted into a diffusion process, after a long-time transition, the critical events reveal their presence through the anomalous asymptotic scaling. This condition, of course, requires that the critical renewal events do not obey Poisson statistics. On the other hand, if the ratio of the number of critical to pseudo events is small, the transition to the anomalous scaling regime might be so slow as to occur outside the range of the numerical analysis of real data~\cite{francesco}.

In this paper we find that the procedure adopted in Ref.~\cite{paradisi} has also the surprising effect of disclosing the hidden power-law tail of the Mittag-Leffler function predicted by the theory of Refs.~\cite{fin1,fin7,fin2}. We propose also a coarse-graining procedure that has the effect of erasing many pseudo events so to make the resulting process almost completely renewal. 

\section{Aging of a non Poisson renewal process}\label{nonpoisson}

To illustrate the basic concepts underlying the method used in this paper, consider the paradigmatic case of a particle moving
with the following equation of motion:
\begin{equation}
\label{dynamicmodel}
  \dot{y} = \alpha y^z ,
\end{equation}
with $y \in [0,1]$,  $z>1$ and $0 < \alpha \ll 1$.
When the particle reaches $y =1$ it is back injected uniformly between $0$ and
$1$ to a totally random initial
position. Thus we define as event the arrival of the particle at the border $y=1$, and we prove with a straightforward algebra that the time distance between two consecutive events yields a waiting time distribution $\psi(\tau)$ with the following form:
\begin{equation}\label{brandnew}
  \psi(\tau) = (\mu - 1) \frac{T^{\mu-1}}{(T + \tau)^{\mu}},
\end{equation}
where $\tau$ is the time it takes for the particle to reach the border,
moving from the initial random position, and:
\begin{equation}
  \mu = \frac{z}{z-1} \qquad , \qquad T = \frac{\mu-1}{\alpha}.
\end{equation}
This is therefore a simple prescription to create a non-Poisson renewal
process. It is also evident that, when $z \to 1$, $\mu \to \infty$, thereby making the waiting time distribution become an exponential function.
  
Let us use the dynamic model of Eq.~(\ref{dynamicmodel}) to illustrate 
the concept of \textit{renewal aging}. For this purpose, let us imagine that the first back injection occurs at time $t=0$, the time origin. We define this condition as \emph{system preparation} at time $t=0$.  We define Eq.~(\ref{brandnew}) as the \textit{brand new
  waiting time distribution}, insofar as it is produced by preparing and observing the system at the same time $t=0$.  
Eq.~(\ref{brandnew}) predicts the time we have to wait to observe the next back injection. If observation begins at time
$t^{\prime} > 0$, and many back injections might have possibly occurred at earlier times, we obtain the waiting time distribution $\psi(\tau,t')$. This waiting time distribution predicts the probability of the first back injection,  with the system 
prepared at $t=0$, and observation beginning at time $t' > 0$.
The function $\psi(\tau,t')$ of Poisson systems is independent of $t^{\prime}$, and in the non-Poisson case it turns out to depend on $t^{\prime}$. An exact
formula for $\psi(\tau,t')$ exists~\cite{pre}, but it is not a simple analytical
expression.  In this paper we shall use instead the following
approximated expression:
\begin{equation}\label{approx}
  \psi(\tau,t') = \frac{\int_0 ^{t'} dy \psi(\tau+y)}{K},
\end{equation} 
where $K$ is a suitable normalization constant. The validity of Eq.~(\ref{approx}) has been already tested for several values
of $\mu$ (see, for instance~\cite{paradisi} for an application on blinking
quantum dots) proving itself to be very reliable.
In the following Sections, instead of the waiting time
distribution $\psi(t, t^{\prime})$, we shall use the SP. The SP is defined as follows:
\begin{equation}
  \Psi(\tau) = \int_t^\infty \psi(y) dy.
\end{equation}
It represents the probability that no events occurred between $0$ and $t$.
The $SP$ corresponding to Eq.~(\ref{approx}) will be denoted by means of the symbol
$\Psi_{t^{\prime}}(\tau)$.

 It is worthwhile spending some more words on  the property of renewal aging, on which the main tenet of this paper rests. Renewal aging is a property of non-Poisson renewal systems that emerges at the level of ensemble. If we consider an ensemble of
trajectories, all of them being different realizations of the same experiment, and
therefore all of them being statistically equivalent, the distribution of the
first back injection times shows renewal aging, that is, a marked dependence on the time at which observation begins. This does not conflict with the fact that at the level of the
single trajectories every event resets to zero the system's memory, thereby producing a condition  where every event is completely uncorrelated with the previous ones.

\section{Illustration of the data under study and empirical analysis of the survival probability}
The data set avaliable to us consists of all the transactions extracted from
the futures on the Italian stock index\footnote{As of 22 March 2004
  the old futures contract on MIB30, called FIB30, has been replaced by the S\&P MIB Futures, because the stock
  index MIB30 has been replaced by the S\&P MIB.}, ranging from January 2000 to
December 2002,  and all the transactions extracted from the futures US stock
index, S$\&$P500,  ranging from January 1993 to January 2001. We have all the
transactions but we use only the next-to-expiration ones, with the markets expiring
quarterly. A futures is a standardized contract that rules the exchange of a
good at a given time, called \emph{maturity}. The futures belongs to a
larger category of contracts, called \emph{derivatives}, that move
nowadays almost $90\%$ of the total amount of money on a market and for this
reason they have become one of the main investigating fields. 

In the
financial markets the price movements are recorded. Thanks to the introduction
of computers, it has become possible to increase the frequency of registration
and therefore explore what is called \emph{high-frequency regime}. In this
paper we focus our attention on this regime. In finance it is common to use,
instead of the price movement, the \emph{return}, defined as the difference of
two consecutive logarithmic prices.

Let us study the distribution of waiting times of the two data sets
avaliable to us. In a recent paper, Mainardi \textit{et. al}~\cite{fin7}
studied the \emph{tick-by-tick} dynamics of the BUND futures traded
at LIFFE. The outcome of their analysis confirmed the statement of a previous
paper by Scalas \textit{et al.}~\cite{fin1} in which the authors argued the
SP to be a Mittag-Leffler function. The Mittag-Leffler function
is an important entire transcendental function~\cite{mittag-leffler}, successfully adopted in
econophysics~\cite{fin1}. It can expressed as a power series in the complex
plane:
\begin{equation}
  E_{\beta}(z) = \sum_{n=0}^{\infty} \frac{z^n}{\Gamma(\beta n + 1)}, \quad \beta >
  0, z \in \Cset ,
\end{equation} 
where $\beta$ is the order of the function. The interested reader can consult,
among the others, Refs.~\cite{fin1,fin7,mittag-leffler} for a complete treatment of the properties of the Mittag-Leffler
function. In a later paper Raberto \textit{et al.}~\cite{fin2} analyzed the statistical properties the SP of General
Electric stock returns and found that it was compatible with a stretched
exponential, which it is known to represent the limit of a Mittag-Leffler
function of argument ($- \gamma t$) and order $\beta$ for $t \to 0$ when $0 < \beta < 1$.

\begin{figure}[!ht]
 \includegraphics[height=12cm,width=6cm,angle=270]{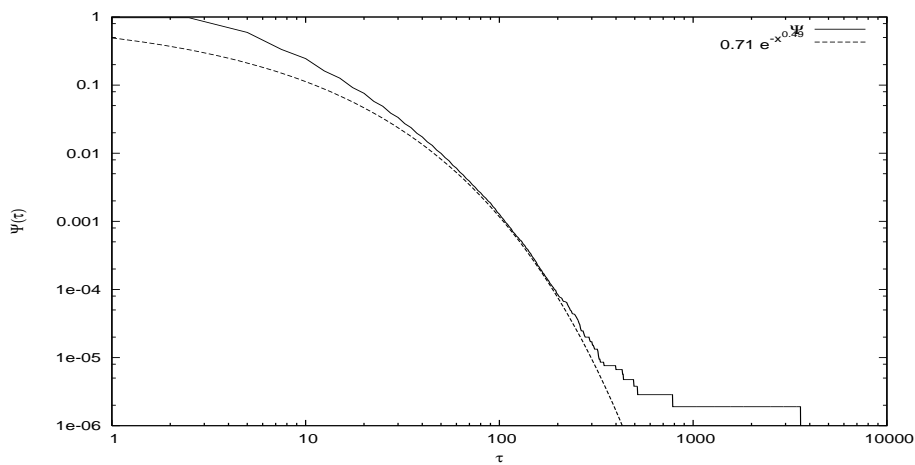}
\caption{In this Figure we show the SP of tick-by-tick returns for the
  Italian S\&P MIB. The fitting function is compatible with a stretched
  exponential $exp\{-(\gamma \tau)^{\beta}/\Gamma(1+\beta)\}$, with $\beta =
0.49$ and $\gamma = 0.4$.}\label{wt_fib}
\end{figure}

\begin{figure}[!ht]
 \includegraphics[height=12cm,width=6cm,angle=270]{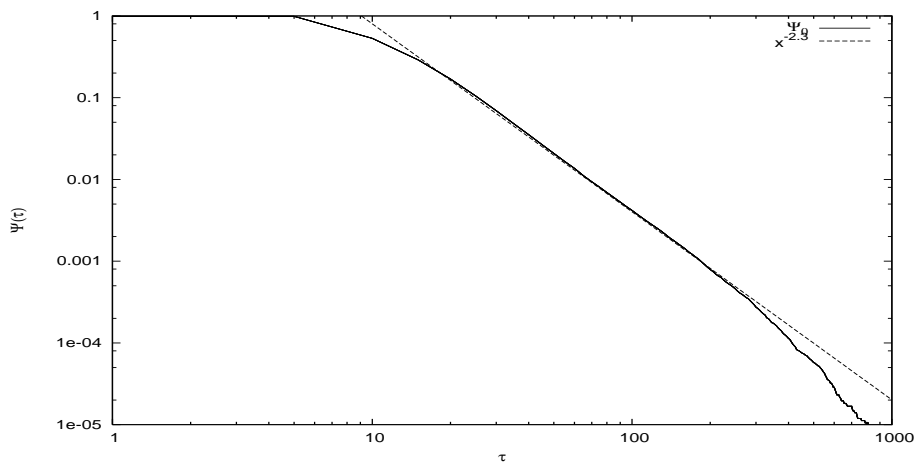}
\caption{In this figure we show the SP of tick-by-tick returns for the US futures on the
  stock index S$\&$P500. The fitting function is compatible with an inverse
  power law function with index $\beta = 2.3$.}\label{wt_us}
\end{figure}

We approach the problem of the statistical analysis of our data set  keeping in mind the interesting results of these authors \cite{fin1,fin7}. In the spirit of these papers, we build the SP 
for the tick-by-tick dynamics of the returns, and we obtain the results of
Figs.~\ref{wt_fib} and~\ref{wt_us}. 

In the case of the Italian futures the best fitting function is a stretched
exponential $exp\{-(\gamma \tau)^{\beta}/\Gamma(1+\beta)\}$, with $\beta =
0.49$ and $\gamma = 0.4$. This result is in
line with  Ref.~\cite{fin2}, where a similar behavior was found for the SP, thereby
implying the absence of the inverse power law tail for the Mittag-Leffler
function. 
In the case of the US futures, Fig.~\ref{wt_us}, we observe instead an inverse power
law function for the SP of returns. The index is $\beta = 2.3$. In this case
we would be tempted  to rule out the theory of the authors of Refs.~\cite{fin1,fin7,fin2}.  In either cases, however,  there are no compelling arguments to prove the CTRW origin of these SP, although in the former
case the result is so close to the theory of Refs.~\cite{fin1,fin7,fin2} as to make it natural to consider it to be a renewal process.  In
the next Section we shall illustrate an aging experiment which will help us to make progresses
towards the important goal of proving the renewal nature of these economic processes.

\section{Aging of the survival probability}

As earlier mentioned, we devote this section to examining our data by means of a procedure applied in a previous article~\cite{paradisi}. This technique turns the SP of the original time series, $\Psi(\tau)$, into two
SP's, denoted as $\Psi_{t^{\prime}}^{exp}(\tau)$ and $\Psi_{t^{\prime}}^{ren}(\tau)$. The former is derived from the original time series by means of an aging numerical experiment that we describe hereby. The latter is derived from the $\psi(\tau)$ of the original time series, as follows. First we turn the experimental $\psi(\tau)$ into $\psi(\tau,t^{\prime})$ by means of Eq.~(\ref{approx}), then we derive from this aged waiting time distribution the corresponding SP, thereby producing $\Psi_{t^{\prime}}^{ren}(\tau)$. In the ideal case of a renewal system we would obtain $\Psi_{t^{\prime}}^{ren}(\tau) = \Psi_{t^{\prime}}^{exp}(\tau)$. According to Ref.~\cite{paradisi}, the existence of pseudo events would reduce the aging effect till to its total annihilation when the number of pseudo events becomes very large. 

The formula of Eq. (\ref{approx}) is based on probabilistic arguments, and consequently, on the ideal use of a Gibbs ensemble of sequences. We have only one sequence, and we have to turn it into a set of Gibbs sequences, prepared in the same way at time $t=0$ and observed at a later time $t^{\prime} > 0$. This is done as follows. The first sequence of the Gibbs ensemble is: $\tau_1^1$, $\tau_2^1$, $\tau_3^1$, $\tau_4^1$ $\ldots$, the superscript indicating the sequence and the subscript the
chronological order of waiting times; the second sequence is obtained from the first by erasing the first waiting time,
thereby yielding $\tau_1^2 = \tau_2^1$, $\tau_2^2 = \tau_3^1$ $\ldots$;
the
third sequence is obtained from the second again erasing the first
waiting time, thus yielding $\tau_1^3 = \tau_2^2 = \tau_3^1$, $\tau_2^3 = \tau_3^2 = \tau_4^1$, and so on. All these sequences are examined in the absolute time representation, where time is denoted by the symbol $t$ and the time of an event occurrence is denoted by the symbol $t_{i}^j$, with the superscript $j$ indicating the system of the ensemble built up with the procedure earlier described.  Note that $t_{1}^j = \tau_{1}^j$, $t_{2}^j = \tau_{1}^j + \tau_{2}^j$,
$t_{i}^j = \tau_{1}^j +\ldots +\tau_{i}^j$. All the systems of this ensemble have been prepared in the same way, with a back injection occurring at time $t = 0$. At this stage we fix the observation time $t^{\prime}$, and for the $j$-th system of the ensemble we monitor the first event occurring after this time, at time $t_{i}^{j} > t^{\prime}$ and the corresponding time distance $\tau_{i}^{j}(t^{\prime}) = t_{i}^{j} - t^{\prime}$. This allows us to define the experimental waiting time distribution $\psi_{t^{\prime}}^{exp}(\tau)$ and from it $\Psi_{t^{\prime}}^{exp}(\tau)$. 
In the case where there is no aging, we should obtain $\Psi_{t^{\prime}}^{exp}(\tau) = \Psi_{t^{\prime} = 0}^{exp}(\tau) = \Psi(\tau)$. The ideal renewal condition should yield $\Psi_{t^{\prime}}^{exp}(\tau) = \Psi^{ren}_{t^{\prime}}(\tau)$.

\begin{figure}[!ht]
 \includegraphics[height=12cm,width=6cm,angle=270]{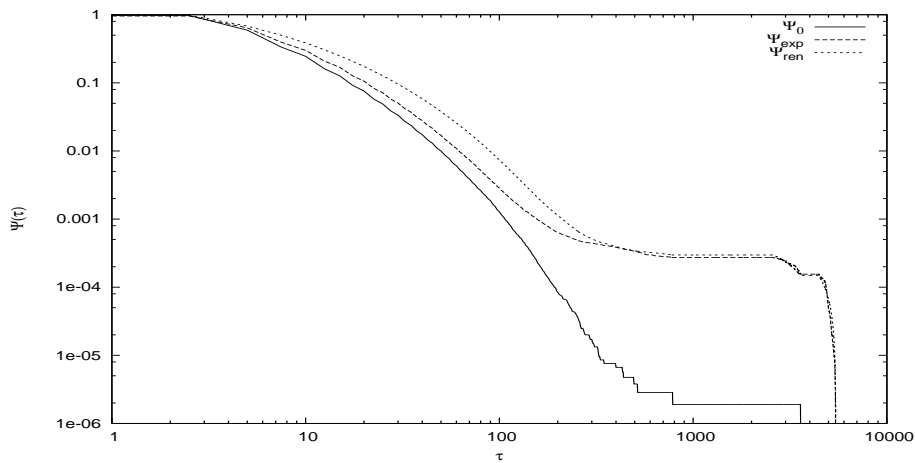}
\caption{In this Figure we report the result of the renewal aging experiment
  carried on the SP of tick-by-tick returns extracted from the futures on the Italian stock
  index. The continuous line represents the non aged SP, called $\Psi_0$, the dashed
  line is the SP evaluated for $t'>0$, $\Psi^{exp}_{t'}$, while the dotted line is the SP
  predicted by the renewal theory, $\Psi^{ren}_{t'}$. $t' = 1000$. We can see aging
  effects. $\Psi^{exp}_{t'}$ is slightly smaller than $\Psi^{ren}_{t'}$ in the first
  part of the plot, while they seem to coincide in the final part.}\label{aging_wt_fib}
\end{figure}

\begin{figure}[!ht]
 \includegraphics[height=12cm,width=6cm,angle=270]{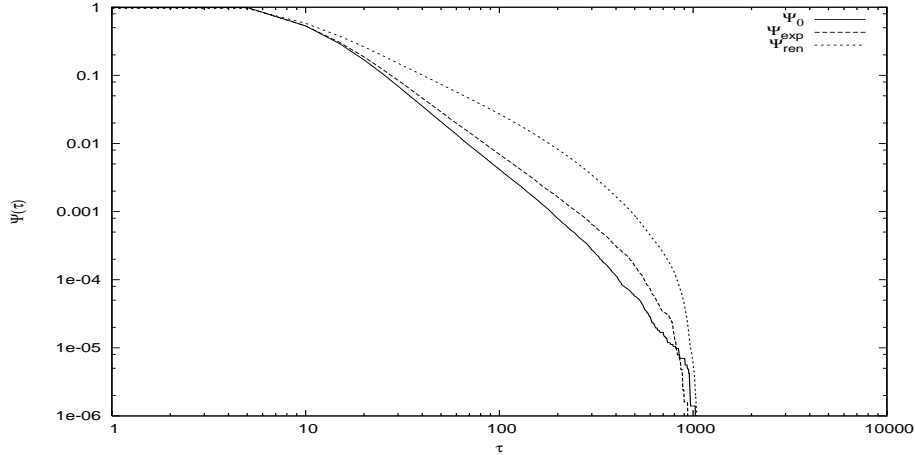}
\caption{In this Figure we report the result of the renewal aging experiment
  carried on the SP of tick-by-tick returns extracted from the futures on the US stock
  index. The continuous line represents the non aged SP, called $\Psi_0$, the dashed
  line is the SP evaluated for $t'>0$, $\Psi^{exp}_{t'}$, while the dotted line is the SP
  predicted by the renewal theory, $\Psi^{ren}_{t'}$. $t' = 1000$. As we can see,
  there are aging effects, but they are strongly reduced with respect to the
  prediction of the renewal theory. }\label{aging_wt_us}
\end{figure}

Fig.~\ref{aging_wt_fib} illustrates the result of this analysis applied to the Italian futures index, for $t'=1000$. Clear aging
effects are evident, slightly reduced in the first part of the plot, where, on the other hand, also the renewal aging is significantly reduced, but
compatible with the predictions of the renewal theory in the second
part. Fig.~\ref{aging_wt_us} reports the same experiment for the US stock
futures index, with $t' = 1000$. In this case, $\Psi^{exp}_{t'}$ is faster than
$\Psi^{ren}_{t'}$ in the whole observed time regime. A weak form of aging is found, but there are no time regions where it correspond to the renewal theory prediction. We believe that the reduced aging is due to microstructure, inducing correlations in the high-frequency returns, see~\cite{simone,simone2,fin2}. It is well known that, for instance, bid-ask spread causes
spurious serial correlations in the time series of returns, with extinction
time of the order of some minutes. The aging experiment is sensitive to
time correlated events (pseudo events), time correlation being a property conflicting with the renewal
assumption of the process, thereby explaining the results of
Fig.~\ref{aging_wt_us}.\\ It is also worth noticing the emergence of a
distinct power law tail in the last part of Fig.~\ref{aging_wt_fib}. This could be a sign of
a hidden Mittag-Leffler-type function as SP, whose presence could be revealed
by means of a renewal aging experiment.  


In conclusion, our analysis proves that the CTRW is a reliable model for the Italian futures index, while in the US futures index the renewal properties are weakened  by 
significant serial correlations in the return time series.

\section{Aging and sign change as a visible event}
\begin{figure}[!ht]
 \includegraphics[height=12cm,width=6cm,angle=270]{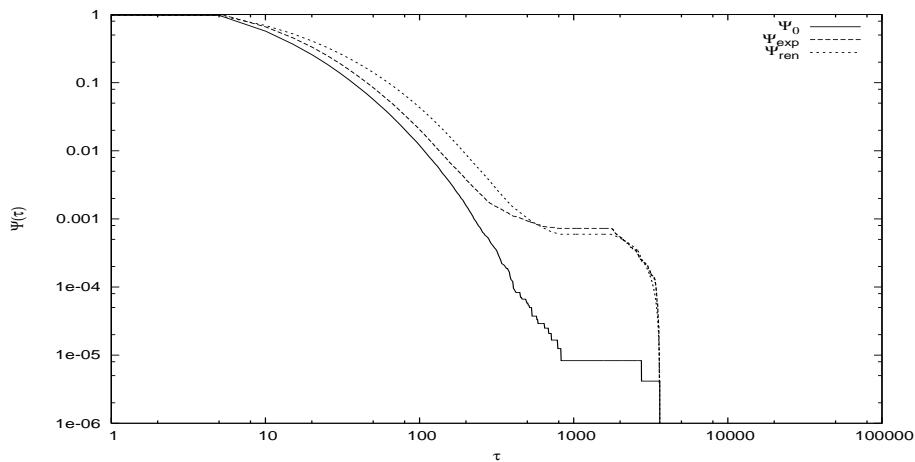}
\caption{In this Figure we report the results of our renewal aging experiment
  carried on the SP of the Italian futures index returns sign persistence. $t' =
  1000$. We can see in this case the presence of aging effects, reduced in the
  first part, while compatible with the renewal assumption in the final part
  of the plot.}\label{pers_fib}
\end{figure}

\begin{figure}[!ht]
 \includegraphics[height=12cm,width=6cm,angle=270]{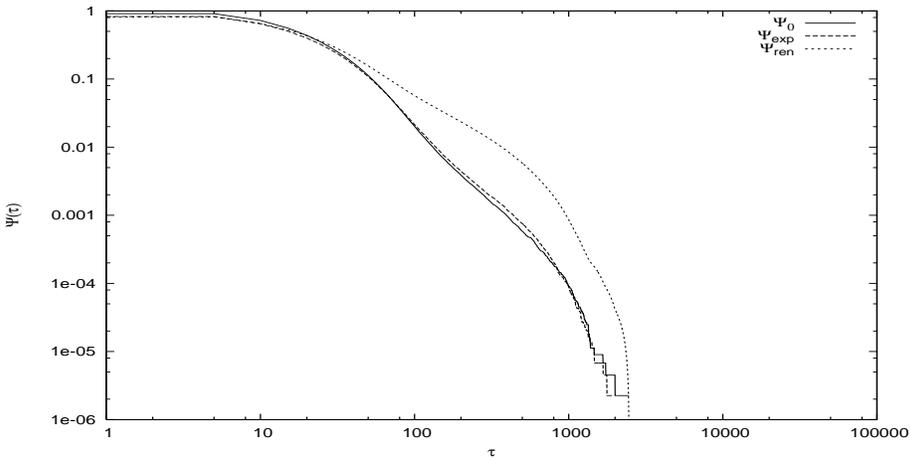}
\caption{In this Figure we report the results of our aging experiment carried on the
  SP of the US futures index returns sign persistence. $t'=1000$. We can see in
  this case the almost total absence of aging effects.}\label{pers_us}
\end{figure}

We move now our attention to a different aspect of the trade action: we study for how long the fluctuation making  the walker (return)  move, maintains the same sign.  Thus, the waiting time $\tau$ is now determined by the time distance between two consecutive changes of sign. We evaluate
the SP of this sequence and study its renewal properties via the renewal aging analysis. Also in this case, we
expect to find reduced aging effects, due to the presence of
negative serial correlations in the time series of returns, with extinction
times of the order of some minutes~\cite{simone}. Figs.~\ref{pers_fib} and~\ref{pers_us} show
the results of our analysis.

As we can see in Fig.~\ref{pers_fib}, in the case of the Italian market, again aging effects are
present. As in the case of the tick-by-tick SP, the aging is reduced and we
can safely impute this behavior to the presence of negative serial correlation
in the time series of returns. However, in the final part of the plot, we have
a good agreement with the renewal assumption for the dynamics of the market.
Regarding the US futures index, Fig.~\ref{pers_us} shows the result in this case. We can see that in this case the aging is almost totally
suppressed. Again we explain this effect invoking the presence of strong
correlations in the time series of returns.

It is important to notice that in the special case where the change of sign of the fluctuation driving the walker's motion is as random as a fair coin tossing, the new waiting time distribution has the same power as the original \cite{zumofen}. We see that Figs.~\ref{pers_fib} and~\ref{pers_us} illustrate a situation that it is qualitatively similar to that of Figs.~\ref{aging_wt_fib} and Figs.~\ref{aging_wt_us}, respectively.  This suggests that the change of sign is a random process, or a process with a modest correlation that might be responsible for the significantly reduced aging effect of the Italian market, which is, in fact, even weaker than in the case of Fig. 4.  

\section{Coarse graining procedure}
We have earlier mentioned that the memory generated by microstructures may hide the renewal nature of the process, which we believe to be the genuine property of the trading action. In the economic literature a coarse graining procedure is usually
applied to raw data to get rid of these memory properties.
Following Ref.~\cite{simone,simone2} we apply the following
tick coarse graining. Let us denote by $[0,T]$ the time window under study, with $T=1$ trading day, that is, 495 minutes, we construct a grid of
$T/\Delta t + 1$  points, with $\Delta t$ denoting the grid spacing.
At every point in time, we define the event (price) as the event (price) of the last transaction before that point. Different
choices, like linear interpolation, are known to introduce spurious
autocorrelation, see, for instance, Ref.~\cite{robbarucci}. However, selecting the length $\Delta t$ of the intervals is conditioned by the following tradeoff. If the interval is too narrow, microstructure effects, such as the bid-ask bounce, remain, which would induce a spurious negative serial correlation. On the other hand, if the interval is too large, there is a loss in the total number of data. 
In order to assess the proper length of the interval we study
the daily variance of the distribution; if the dynamics of the process is a
stochastic differential equation with no drift, then the integrated variance
on the time  window $T$ must coincide with its expected value for every value
of the grid spacing $\Delta t$, see, for instance,
Ref.~\cite{SDE}.

\begin{figure}[!ht]
 \includegraphics[height=6cm,width=12cm]{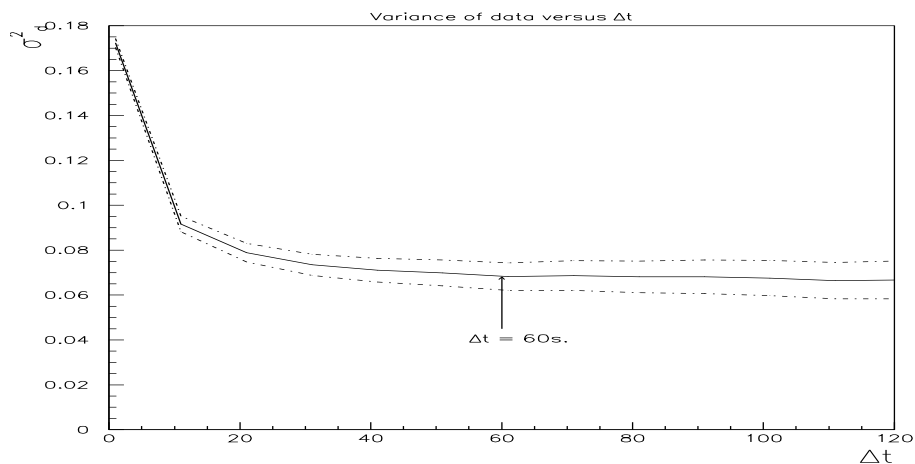}
\caption{In this Figure we show the integrated variance of the distribution of returns for the
  Italian futures stock index S$\&$P FIB as a function of the grid spacing
  $\Delta t$.}\label{varianza}
\end{figure}

These important issues are illustrated by Fig.~\ref{varianza}, which shows the integrated variance of the
Italian futures index versus the length of the interval. In accordance with the earlier remarks, we see that, when the spacing is too narrow,
the variance increases. This fact is well known and can be imputed to the
presence of microstructure effects~\cite{roll}. We decide to use
an interval spacing of $60$ seconds for the Italian futures, and $240$ seconds
for the US futures, where we assume the integrated daily variance to be
constant.  Indeed, the percentage of empty intervals in the two case
has been estimated, respectively, to be below $1\%$ and below $3\%$.

We are now properly equipped to establish if the nature of the process is compatible with the
renewal assumption. The discussion of the results emerging from the application of this procedure will be the argument of the next Section.

\section{Pseudo events, critical events and renewal aging }
The adoption of the coarse graining procedure described in the previous section should make it possible to confirm our conviction that the weakening of the aging effects depends on pseudo events.  As already stated above, the presence of
microstructure effects implies high-frequency spurious correlation effects in
the time series of returns, that is, the real event is ``floating'' in a sea
of pseudo events. We argue that, if we were able to distinguish the real event from the
pseudo events,  and create the waiting time distribution of the real events, the resulting 
distribution, if markedly  non exponential, should obey the condition of renewal aging. The coarse graining has the effect of reducing the number of pseudo events,
thereby  disclosing the true nature of the
financial market. 

Fig.~\ref{aging_fib} and Fig.~\ref{aging_sp} show the results of the aging experiment on the Italian
futures index and on the US futures index, respectively. We see that in both cases the aging effect is very close to the predictions of the renewal theory. It is worth noticing that according to Fig.~\ref{aging_fib}, the aging
experiment in the long time region makes, again, an inverse power law tail emerge for both $\Psi_{t'}^{exp}$ and $\Psi_{t'}^{ren}$. The shape of the function
becomes a Mittag-Leffler type with $\beta = 0.95$ and $\gamma = 0.5$. This
behavior remains unchanged for larger values of $t'$, therefore suggesting
that the aging experiment reveals the genuine nature of the process,
this being represented by the Mittag-Leffler SP found by the authors of
Ref.~\cite{fin7}. We think that this may become an efficient way to assess the
hidden renewal nature of the experimental time series and we plan to devote
further study to this issue in a later publication. \\
\indent The fact that the aging effect in the US market is weaker than in the Italian market is not due to the non-renewal nature of the US market, but it seems to depend on the fact that the Mittag-Leffler form of the SP is lacking in this case. We cannot rule out  the possibility that also the US market obeys the theory of Ref.~\cite{fin1}. In fact, our results suggest that in this case the stretched-exponential regime of the US market is more extended than that of the Italian market. We note that in the short-time regime of Fig.~\ref{aging_fib}, namely, in the stretched-exponential time region, aging is virtually absent, in spite of the renewal character of this process. Thus, the reduced aging cannot be used as a sign of departure from the renewal condition. We think that the more significant aging effect in the large time regime of Fig.~\ref{aging_sp} is a sign that also in this case the inverse power law of the Mittag-Leffler function is made to emerge by the aging experiment.

\begin{figure}[!ht]
  \includegraphics[width=6cm, height=12cm, angle=270]{./aging_fib.epsi}
  \caption{In this Figure we report the aging experiment for the Italian futures market, S$\&$P
  MIB, after the coarse graining procedure ($\Delta t = 60 sec.$).  The continuous line  represents the non aged
  SP $\Psi_0$, the dashed line is the experimental SP $\Psi^{exp}_{t'}$,
  while the dotted line  is the SP predicted by the renewal theory
  $\Psi^{ren}_{t'}$. $t^{\prime} = 1000$. In this case the system seems to age
  compatibly with the renewal assumption.}\label{aging_fib}
\end{figure}

\begin{figure}[!ht]
  \includegraphics[width=6cm,height=12cm, angle=270]{./aging_sp.epsi}
\caption{In this Figure we report the aging experiment for the futures on US the stock index, S$\&$P
  500, after the coarse graining procedure ($\Delta t = 240 sec.$). The continuous line  represents the non aged
  SP $\Psi_0$, the dashed line is the experimental SP $\Psi^{exp}_{t'}$,
  while the dotted line  is the SP predicted by the renewal theory
  $\Psi^{ren}_{t'}$. $t^{\prime} = 1000$. Also in this case the system seems
  to age in a way compatible with the renewal assumption. The aging is however
  reduced with respect to the previous case.}\label{aging_sp}
\end{figure}

\section {Concluding remarks}
This paper confirms the theory proposed by the authors of Refs.~\cite{fin1,fin7,fin2} to describe the financial market dynamics, with a careful scrutiny of the economic data. The Mittag-Leffler SP, which is the main  prediction of this theory,  does not emerge in its entirety from the data analyzed in this paper. The statistical inaccuracy of the long-time region 
obscures the emergence of the inverse power law tail in the US financial
market, and the Italian financial market as well. The aging experiment has the
surprising effect of making the inverse power law distinctly appear in the
long-time regime of the Italian market. The effect is not so evident in the
case of the US market, and we think that this is due to fact that the
stretched-exponential regime is much more extended than in the case of the
Italian market. It is remarkable that the genuinely renewal nature of the
trading action is made visible, through the aging experiment, by a proper
coarse graining data processing. Our analysis suggests also that the price persistence is not a process with long-range memory, and within the limits of the numerical accuracy, the price change is virtually indistinguishable from the adoption  of a fair coin tossing to decide if price has to increase or to decrease. 

We hope that the results of this paper may give an important contribution to the foundation of accurate financial models. Furthermore, we are convinced that the procedure illustrated in this article is a significant contribution to the detection of invisible renewal events, the challenge emerging from the results of Ref.~\cite{memory}.

\ack{
\noindent We thankfully acknowledge Dr. R. Ren\`o for providing
the excellent S$\&$P FIB data set, as well as Borsa Italiana SPA, and in particular
Concetta Ricciardi and Ada De Roma. We also thankfully acknowledge Welch and ARO for financial support through Grant no. B-1577 and no. W911NF-05-1-0205, respectively.}


\begin{thebibliography}{100}

\bibitem{fractional} E. Scalas, Physica A, {\bf 362}, 225-239 (2006).
\bibitem{ctrw} E.W. Montroll, G.H. Weiss,  J. Math. Phys. {\bf 6}, 167 (1965). 
\bibitem{kenkre} V.M. Kenkre, E.W. Montroll, and M.F. Shlesinger, J. Stat. Phys. {\bf 9}, 45 (1973). 
\bibitem{balescu} R. Balescu, Chaos, Solitons and Fractals,
{\bf xx}, www (2006), this issue. 
\bibitem{rasit} R. Cakir, A. Krockin, P. Grigolini, {\bf xx}, www (2006), this issue. 


\bibitem{fin1} E. Scalas, R. Gorenflo, F. Mainardi, Physica A, {\bf 284}, 376 (2000). 
\bibitem{fin7} F. Mainardi, M. Raberto, R. Gorenflo, E. Scalas, Physica A,
  {\bf 287}, 468-481 (2000).

\bibitem{fin2} M. Raberto, E. Scalas, F. Mainardi, Physica A, {\bf 314}, 749 (2002).

\bibitem{fin3} M. Kirane, Y. Laskri, N. -e. Tatar, J. Math. Anal. Appl. {\bf 312}, 488 (2005). 

\bibitem{fin4} T. Kaizoji, M. Kaizoji, Physica A, {\bf 336}, 563 (2005).

\bibitem{fin5} L. Palatella, J. Perell\'{o}, M. Montero, J. Masoliver,
Physica A, {\bf 355}, 131 (2005).

\bibitem{fin6} S. -M. Yoon, J.S. Choi, Y. Kim, K. Kim, 
Physica A, {\bf 359}, 131 (2006).

\bibitem{grigo} P. Grigolini, Adv. Chem. Phys. , in press (2006). 

\bibitem{barkaiaging} E. Barkai and Y.-C. Cheng, J. Chem. Phys. {\bf 118}, 6167 (2003). 

\bibitem{brok} X. Brokmann, J. -P. Hermier, G. Messin, P. Desbiolles,
  J. -P. Bouchaud, and M. Dahan, Phys. Rev. Lett. {\bf 90}, 120601 (2003). 

\bibitem{paradisi} S. Bianco. P. Grigolini, P. Paradisi, J. Chem. Phys. {\bf 123}, 174704 (2005). 

\bibitem {memory} P. Allegrini, P. Grigolini, P. Hamilton, L. Palatella, and
  G. Raffaelli, Phys. Rev. E {\bf 65}, 041926 (2002).

\bibitem{mega} M. S. Mega, P. Allegrini, P. Grigolini, V. Latora, L. Palatella, Phys. Rev. {\bf 90}, 188501 (2003).  

\bibitem{francesco} P. Allegrini, F. Barbi, P. Grigolini, submitted to Phys. Rev. E. 


\bibitem{mittag-leffler} F. Mainardi and R. Gorenflo, J. Comput. and Appl. Mathematics {\bf 118}, 283 (2000). 



\bibitem{pre} P. Allegrini, G. Aquino, P. Grigolini, L. Palatella, A. Rosa,
  and B.J. West, Phys. Rev. E {\bf 71}, 066109 (2005). 

\bibitem{simone} S. Bianco and R. Ren\`o, Journal of Futures Market, {\bf 26}(1),
  61-84 (2006).

\bibitem{simone2} S. Bianco and R. Ren\`o, in preparation.

\bibitem{zumofen} G. Zumofen and J. Klafter
Phys. Rev. E 47, 851-863 (1993).

\bibitem{robbarucci} E. Barucci and R. Ren\`o, Econ. Lett., {\bf 74}, 371-378 (2002).

\bibitem{SDE} O.E. Barndorff-Nielsen and N. Shephard, Journal of the Royal
  Statistical Society, Series B, {\bf 64}, 253-280 (2002).

\bibitem{roll} R. Roll, Journal of Finance, {\bf 39}(4), 1127-1139 (1984).

\end{thebibliography}
\end{document}